\documentclass[a4paper,english]{article}
\usepackage{pslatex}
\usepackage[T1]{fontenc}
\usepackage[latin1]{inputenc}
\usepackage[authoryear]{natbib}

\makeatletter

\newenvironment{lyxcode}
{\begin{list}{}{
\setlength{\rightmargin}{\leftmargin}
\setlength{\listparindent}{0pt}
\raggedright
\setlength{\itemsep}{0pt}
\setlength{\parsep}{0pt}
\normalfont\ttfamily}%
 \item[]}
{\end{list}}

\usepackage{babel}
\makeatother
\begin{document}

\title{Escalating The War On SPAM Through Practical POW Exchange}

\author{Paul Gardner-Stephen (\texttt{paul@infoeng.flinders.edu.au}),\\
School of Informatics \& Engineering,\\
Flinders University, Adelaide, Australia}

\maketitle
\begin{abstract}
Proof-of-work (POW) schemes have been proposed in the past. One prominent
system is HASHCASH \citep{back-hash} which uses cryptographic puzzles
. However, work by \citet{laurie04proofwork} has shown that for a
uniform proof-of-work scheme on email to have an impact on SPAM, it
would also be onerous enough to impact on senders of {}``legitimate''
email. I suggest that a non-uniform proof-of-work scheme on email
may be a solution to this problem, and describe a framework that has
the potential to limit SPAM, without unduly penalising legitimate
senders, and is constructed using only current SPAM filter technology,
and a small change to the SMTP (Simple Mail Transfer Protocol). Specifically,
I argue that it is possible to make sending SPAM 1,000 times more
expensive than sending {}``legitimate'' email (so called HAM). Also,
unlike the system proposed by \citet{liu06}, it does not require
the complications of maintaining a reputation system. 
\end{abstract}

\section{Introduction}

\subsection{A Brief Introduction To Proof-Of-Work As A SPAM Counter Measure}

Proof-of-Work (POW) systems were suggested as a counter measure to
junk email (SPAM) as early as 1992 \citep{705669}, and the concept
has been rediscovered and developed since then, e.g, \citep{rivest96payword,888615,jakobsson99proofs,back-hash,dwork02memorybound,830561,1064341,DBLP:conf/acns/2005}.

In simple terms a POW system uses a puzzle challenge that is hard
to solve, but easy to verify. The solution to the puzzle is presented
to the receiving email server as proof of having completed a certain
amount of work. The intention is that this will limit the number of
messages that a Spammer can send per unit time. When applied to all
email messages, I call the scheme a \emph{Uniform Cost Proof Of Work}
Scheme. That is, the computational burden is equal for every message,
whether it is wanted by the recipient (HAM), or unwanted by the recipient
(SPAM). 

In reality, due to differences in computational capacity (CPU speed)
of mail servers, the burden is not uniform. These differences can
be great, probably around two orders of magnitude between the fastest
and slowest computers that send email. However, the difference between
the fastest and slowest random access memories are much less than
the difference between the fastest and slowest CPUs. Classes of puzzles
have been suggested that use this property to keep the burden as uniform
as possible, to reduce the performance difference to a factor of less
than five \citet{1064341}. This would seem to be a reasonable achievement.

\subsection{Problems With Uniform-Cost Proof-Of-Work As A SPAM Counter Measure}

Assuming for the moment, then, that (approximately) Uniform-Cost Proof-of-Work
systems exist, we turn to the problems they face. The following is
a list of problems with Uniform-Cost Proof-of-Work systems that, with
the exception of message latency, are drawn from the work of \citet{laurie04proofwork}
and \citet{securityfocus_pow}:

\subsubsection{Message Latency}

By applying a computational delay to every message, the implied real-time
delivery semantic of email is broken. This makes a Uniform-Cost Proof-of-Work
scheme socially unattractive.

\subsubsection{Inequitable Burden}

The problem of inequitable burden of proof-of-work schemes due to
server speed has already been mentioned, and even though \citet{1064341}
showed how to reduce the margin to less than five-fold, that is still
a significant inequality.

\subsubsection{Mailing Lists}

Mailing lists send a great many messages. If each recipient invoked
a uniform email proof-of-work burden, then large mailing lists would
become expensive, if not impractical. That is, the indiscriminate
nature of the proof-of-work scheme causes mailing lists to be impacted
to the same degree as a Spammer who sends the same quantity of messages.

\subsubsection{Robot Armies}

Moreover, because many Spammer control hundreds of thousands of compromised
computers around the world, it would seem that Spammer have access
to much larger CPU resources than most legitimate senders. Thus a
uniformly applied proof-of-work scheme on email may actually hurt
legitimate senders more than it hurts Spammer. Also, Spammer are able
to maintain legitimate CPU resources, further compounding the problem.

\subsubsection{Summary}

To summarise, we discover what \citet{laurie04proofwork} have established
from a theoretical perspective, that for any uniform email proof-of-work
scheme to be sufficient to reduce SPAM (assuming 2004 levels of roughly
equal SPAM and HAM), then the several percent of legitimate email
senders who send relatively many email messages will be prevented
from doing so. In other words, a Uniform-Cost Proof-of-Work scheme,
regardless of the quality of the algorithms it uses, cannot succeed,
precisely because it burdens SPAM and HAM uniformly. Therefore, if
Proof-of-Work is to work, it cannot be with a Uniform-Cost model.

\subsection{Introducing Targeted-Cost Proof-Of-Work As A Feasible SPAM Counter
Measure}

Having argued that a Uniform-Cost Proof-of-Work scheme cannot succeed,
I suggest that Proof-of-Work schemes, generally, are not fatally flawed.
Note that for every objection in the previous text, that the problem
is the indiscriminate application of a uniform burden on all email
messages. The logical alternative is to apply the burden on a more
intelligent basis. An ideal implementation would place no burden on
legitimate messages (HAM), and an infinite and unavoidable burden
on SPAM. Such a system would immediately, by definition, stop all
SPAM --- unfortunately this result cannot be produced with current
technology.

While we have no 100\% accurate method to discriminate HAM and SPAM,
we do have effective heuristic email classification systems, such
as SpamAssassin%
\footnote{HTTP://spamassassin.apache.org/%
}, CRM114%
\footnote{HTTP://crm114.sourceforge.net/%
} and many others. Thus, by using the judgements of an existing SPAM
filter, as to the probability of a given message being SPAM, it is
possible to dynamically determine the burden to apply to that message.
That is, we can still place the vast majority of the burden on Spammers,
with the precise proportion determined by the quality of the SPAM
filters. As with previous Proof-of-Work schemes, the work would take
the form of solving a specially created cryptographic puzzle with
the desired degree of difficulty.

Such a scheme has a subtle but important difference to just using
SPAM filters alone: The failure mode for falsely classifying HAM as
SPAM is more robust, as the delivery of an incorrectly classified
message is only resisted, instead of being refused. For example, it
would be unrealistic to configure a SPAM filter to reject messages
that are only 5\% likely to be SPAM, even though that may be the threshold
required to reject practically all SPAM. However, it is completely
reasonable to resist the delivery of messages that are 5\% likely
to be SPAM. I call this method of the selective application of a Proof-of-Work
scheme Targeted-Cost Proof-of-Work.

This assumes that all computers can perform the work required by a
Proof-of-Work scheme at a uniform rate. But computers vary in speed.
Fortunately, if both small hand held devices and super-computers are
excluded from consideration (this seems reasonable, since few legitimate
mail senders use hand held devices, and few spammers have sustained
access to super computers), most computers connected to the Internet
vary in clock speed by no more than about one order of magnitude.
Also, research has been conducted into creating problems that are
limited by a computers random access memory speed, which is a quantity
that varies much less than CPU speed \citep{dwork02memorybound,1064341}.
Moreover, there seems to be no reason to suggest that the average
computer being used to send SPAM will be much faster than the average
computer being used to send legitimate email. Thus, the system

Some SPAM filters have demonstrated >99.9\% accuracy%
\footnote{HTTP://crm114.sourceforge.net/%
}. Therefore, by using SPAM filters to inform a Targeted-Cost Proof-of-Work
scheme, it should be possible to correctly resist 99.9\% of SPAM,
but only resist 0.1\% of HAM. Thus Spammers are burdened almost 1,000
times more per message, on average, than legitimate senders. 

By shifting the vast majority of the computation burden from legitimate
senders to Spammers, Proof-of-Work should be feasible. Consider the
calculations presented by \citet{laurie04proofwork} that showed how
a Uniform-Cost Proof-of-Work scheme would hurt legitimate senders
who send more than 250 emails per day. The graph presented by \citet{laurie04proofwork}
suggests that 1.56\% of legitimate senders would fall into this category.
If the cost to send (on average) were reduced by a factor of 1,000,
then legitimate senders could send of the order of 250,000 messages
per computer per day, while still limiting Spammers to of the order
of 250 messages per computer per day. 

Moreover, if the advantage is 1,000 times, then some legitimate sending
capacity can be sacrificed to further limit the sending of SPAM. I
suggest that a Targeted-Cost Proof-of-Work scheme require proof of
approximately 1 hour of work for messages suspected of being SPAM
(assuming 99.9\% accurate classification). In that case, legitimate
senders could deliver about 24,000 messages per computer per day,
while Spammers would be limited to only 24 messages per computer per
day. Assuming that Spammers have about 10 million computers at their
disposal, this would limit daily spam volumes to only 24 million messages
per day --- about one SPAM for every twenty Internet users. 

If the system is detuned to accommodate a SPAM classifier that is
less accurate, then the advantage would be reduced, either by limiting
the daily capacity of legitimate senders, or by allowing Spammers
to send more SPAM, or perhaps a combination of the two. 

Even a classification accuracy of only 95\% gives a 20:1 advantage
to HAM over SPAM, which would allow legitimate senders to deliver
several thousand messages per computer per day, while limiting Spammers
to only a couple of hundred messages per computer per day. These limits
should accommodate practically all legitimate senders (\citet{laurie04proofwork}
suggest that fewer than 0.1\% of users send more than a couple of
thousand email messages per day), while limiting SPAM to similar volumes
as HAM (since the vast majority of legitimate senders send many fewer
messages per day (according to \citet{laurie04proofwork}, the median
is about 90 messages per day).

On balance, requiring about an hour of work for each suspected SPAM
seems reasonable, as it severely limits SPAM delivery capacity, while
not burdening legitimate senders too much: If SPAM filter accuracy
of 99.9\% is sustainable, then a cost of 1 hour applies to only 0.1\%
of legitimate email, giving an average delivery cost of only 3.6 seconds.

Moreover, as will be described below; a) legitimate bulk senders can
act to reduce their burden; and b) it is possible for each receiving
mail server to configure this behaviour independently, reflecting
the accuracy of their SPAM filters, or to satisfy local policy. That
is, a Targeted-Cost Proof-of-Work scheme can improve the HAM to SPAM
advantage somewhat beyond what is immediately apparent. However, there
are risks that must be managed.

\section{Managing Risks}

\subsection{Variable SPAM Filter Performance}

The effectiveness of a Targeted-Cost Proof-of-Work system is proportional
to the accuracy of the SPAM filter that underlies it. But SPAM filters
differ in accuracy from site to site. Some of this variability could
be fixed if all sites used the current best SPAM filter software.
However, some variability is due to the kind of email that sites receive
(consider the difference between what staff at a marketing company,
a SPAM researcher, and staff at a medical centre would consider SPAM
and HAM, and also the difference between the kinds of messages they
might receive).

Fortunately, a Targeted-Cost Proof-of-Work system need not consider
these issues. This is because the receiving mail server specifies
whether to resist delivery of a message, and what the level of resistance
is. The local message administrator can tune the threshold at which
the resistance applies, and the level of the resistance (perhaps introducing
a sliding resistance scale based on the spamminess of a message, and
white lists of senders who are never resisted). Therefore each mail
server can optimise the system to minimise the amount of SPAM that
is delivered, without unduly burdening legitimate senders.

\subsection{When Delivery Of Legitimate Mail Is Too Expensive}

This leads to an important issue: What happens when a receiving mail
server imposes a burden that a legitimate sender is not willing to
meet? This is the one failure mode that is undesirable (SPAM that
is accepted is less of a problem, because the receiving mail server
may still discard the message or mark it as SPAM, even after it has
been accepted from the sender). Ideally, the sending mail server should
alert the user that their message was not delivered because it was
too spammy. The user could then re-draft and re-send the message.
Alternatively, they could send a fresh message to the recipient asking
to be added to their receiving mail servers white list so that future
messages will not be resisted.

\subsection{Senders Of {}``Pressed HAM'' (PHAM) }

A related problem is legitimate senders who send messages that are,
perhaps unavoidably, spammy. I call such messages Pressed HAM (or
PHAM for short) --- they are not quite SPAM, but like Pressed HAM,
they are not as palatable as real HAM. Many solicited advertisements
and business newsletters would fall into this category. First, it
is observed that business senders are the group most able to meet
the delivery burden, assuming that they have a sound and profitable
business model. Secondly, the resistance to delivery creates an economic
incentive for such senders to avoid some of the evils of PHAM. Much
PHAM can be rendered more palatable by including a link to the information,
rather than hundreds of kilobytes of HTML and images.

\subsection{Mailing Lists}

Mailing lists are relatively straight forward: A correctly configured
mailing list, that carries HAM, will not be greatly penalised. In
contrast, if a Spammer uses a mailing list to attempt to deliver SPAM
to a broad audience, then resistance will be applied to most deliveries.
To alleviate this, and to reduce resource requirements, a mailing
list could refuse to deliver messages that invoke too much resistance
during delivery (and alert the sender of this). This may seem severe,
however, some existing mailing list providers already implement similarly
harsh policies, e.g., never re-trying delivery after a temporary failure.

\subsection{Leaking SPAM Filter Information To Spammers}

Because the resistance is selectively applied to SPAM, it is possible
that Spammers may try submitting successive refinements of their messages
to a given mail server in order to try to reduce the spamminess of
the message, and so avoid encountering resistance during delivery.
To mitigate this risk, it is recommended that in practise a mail server
use a single degree of resistance: Either delivery is resisted, or
it is not. If graduated resistance is absolutely required, the more
coarsely grained the graduation, the better. The level of resistance
could include a random factor to make it more difficult to reverse
engineer HAM flavoured SPAM. Having had said this, a finely graduated
resistance scheme makes a lot of sense, in that it encourages senders
to reduce the overall spamminess of their messages, and provides a
more graceful failure mode for misclassification of messages. Therefore
we briefly consider the likely impact of using a graduated resistance
scheme.

Regardless of the resistance scheme that is applied, the impact of
any leaked information about the SPAM filter must be considered. There
are two possible scenarios: 1) There are few distinct SPAM filters;
or 2) There are many distinct SPAM filters. 

If there are few distinct SPAM filters, then the information leaked
by mail servers is already available to Spammers, because it is feasible
for them to run their message through each and (much more easily)
minimise the spamminess of their message. Alternatively, there are
too many SPAM filters for a Spammer to do this. In that case, it seems
reasonable to assume that minimising the spamminess of a message against
any one filter will be of only limited value, as other filters will
target different message features. Moreover, there are characteristics
of a message that a Spammer cannot readily change, such as the IP
address they are sending from. Thus the task of reducing the spamminess
of a given message sufficient to avoid widespread classification as
SPAM is particularly difficult. 

However, let us be pessimistic, and assume that a Spammer does come
up with a process that enables them, by repeated submission of a message,
reduce its spamminess sufficient to avoid resistance during delivery.
In that situation it is possible to use a {}``sin bin'', that blocks
delivery when this anti-social behaviour is observed, e.g., repeatedly
declining to meet the delivery burden imposed on a message. Hosts
in the sin bin may not submit any more mail until a timeout elapses%
\footnote{In theory, a particularly well organised Spammer could use an army
of robots to send each successive refinement from a different IP address.
However, the complexity of this arrangement, combined with the further
increase in bandwidth requirements in order for the robots to coordinate
their actions, would limit the value of this approach.%
}. 

However, even if it were possible to minimise the spamminess of a
message by repeated submission, the repeated delivery attempts would
multiply the network bandwidth required to deliver each message, thus
reducing the amount of SPAM that can be send per unit time. Therefore,
the effects of any information leaked via a Non-Uniform Proof-of-Work
scheme are mitigated at multiple levels.

\section{Implementing The SPAM Friction}

\subsection{Integrating SPAM Friction Into The SMTP}

A Targeted-Cost Proof-of-Work system must see the the body of an email
before it can decide whether to charge for delivery. For this reason,
it is not possible to incorporate it into the existing SMTP \citep{rfc2821}.
Specifically, while a receiving mail server could use a 211 message
to notify the sender that they must provide Proof-of-Work, there is
no mechanism in SMTP that would allow the sender to deliver the Proof-of-Work
receipt. Thus a new keyword would be required in SMTP, or an old one
must take on a new meaning. The former seems preferable, since if
the SMTP must be changed, then it should not be done crudely. I suggest
the new keyword POW as the new keyword that is used to deliver a Proof-of-Work
receipt to the receiving mail server. The following shows an example
conversation using the new keyword:

\begin{lyxcode}
250~ESMTP~Server~Ready

EHLO~sending-mail.com

250-receiving-mail.com~Hello~sending-mail.com~{[}10.1.2.3]

250-SIZE~52428800

250-AUTH~PLAIN~LOGIN

250-STARTTLS

250-SPAMFRICTION~ALG0,~ALG1,~ALG2

250~HELP

POW~ISUPPORT~ALG0,~ALG1,~ALG4

250~OK

MAIL~FROM:~sender@sending-mail.com

250~OK

RCPT~TO:~receiver@receiving-mail.com

250~Accepted

DATA

354~Enter~message,~ending~with~{}``.''~on~a~line~by~itself

Spam,~spam,~eggs~and~spam

.

211~POW~Required~(SPAM)~0:21:892734982734987

POW~RECEIPT~0:21:892734982734987:193287436879263

250~OK~id=1HCVqn-000436-HX
\end{lyxcode}
The POW command is issued in response to a 211 message that requests
proof of work. The string \texttt{0:21:892734982734987} is the puzzle
that the sender must solve, where \texttt{0} is the algorithm number,
\texttt{21} is the difficulty, and the long string of numbers the
remainder of the problem specification. The response contains the
puzzle concatenated with the solution strings. Otherwise, the conversation
conforms to the SMTP.

\subsection{Transitional Considerations}

A critical issue for any changes to the SMTP, is that backward compatibility
is retained. This is why the POW capability is issued using a 250
message, and the receiver is required to acknowledge if it also supports
the capability. A legacy sender that does not support the capability
makes no offer, alerting the receiver of the fact by its silence.
Legacy senders would be handled by introducing a lengthly delay before
offering to accept a message. To prevent abuse by spammers, a receiving
mail server may take several courses of action when waiting for a
Proof-of-Work receipt, or during the time out period when communicating
with a legacy sender: a) refuse to accept additional connections from
the same sending host; b) temporarily reject messages sent from the
same sending host; c) accept additional messages, but increasing the
delivery resistance in accordance with the number of messages in this
state. In any case, it is possible to prevent Spammers from abusing
the time out too much. More to the point, the time out still makes
it harder to send SPAM than is currently the case.

Finally, for the scheme to be fully effective it would require participation
by the majority of SMTP servers on the Internet. The interoperability
with the existing SMTP just described helps to make it possible to
progressively implement the scheme, with value increasing with each
SMTP server that adopts it. Thus if the protocol was implemented in
the major SMTP server software packages, then wide spread adoption
would be possible.

\section{Conclusions}

In this paper I have argued that a Targeted-Cost Proof-of-Work scheme
has the potential to dramatically reduce SPAM volumes. Moreover, its
requirements are modest: a) SPAM filters that are $\ge99.9$\% accurate
(which already exist); b) relatively small changes to the SMTP and
mail server software and configurations. The risk-benefit ratio is
compelling: The chance to make SPAM 1,000 times harder to send than
HAM, and limiting overall SPAM volumes to around 24 messages per sending
computer per day. The obvious next step is to implement this system,
including selecting at least one initial problem class, and assess
its effectiveness.

\section{Acknowledgements}

I would like to acknowledge the valuable feedback and encouragement
of the my colleagues in the School of Informatics \& Engineering and
the LabF community.

\bibliographystyle{plainnat}
\addcontentsline{toc}{section}{\refname}\bibliography{/home/paul/Documents/Papers/SpamTax/spamtax,/home/paul/Documents/Papers/SpamTax/rfc}

\end{document}